\documentclass[conference]{IEEEtran}
\IEEEoverridecommandlockouts
\usepackage{cite}
\usepackage{amsmath,amssymb,amsfonts}
 \usepackage{verbatim}
\usepackage{graphicx}
 \usepackage{url} 
\usepackage{mdframed}
\usepackage{multirow}
\usepackage{booktabs}
\usepackage[table,xcdraw]{xcolor}
\usepackage{textcomp}
\usepackage{xcolor}
\usepackage{algpseudocode}
\usepackage{algorithm}
\usepackage[table]{xcolor} 
\usepackage{multirow}      
\usepackage{booktabs}      
\usepackage[utf8]{inputenc}
\usepackage{tcolorbox}
\tcbuselibrary{skins}

\def\BibTeX{{\rm B\kern-.05em{\sc i\kern-.025em b}\kern-.08em
    T\kern-.1667em\lower.7ex\hbox{E}\kern-.125emX}}
\begin{document}

\title{Automated Research Article Classification and Recommendation Using NLP and Machine Learning}
\author{
  \IEEEauthorblockN{
    Shadikur Rahman\textsuperscript{*}\textsuperscript{†},
    Hasibul Karim Shanto\textsuperscript{‡},
    Umme Ayman Koana\textsuperscript{*}\textsuperscript{†},
    Syed Muhammad Danish\textsuperscript{†}
  }
  \IEEEauthorblockA{\textsuperscript{*} York University, North York, Canada}
  \IEEEauthorblockA{\textsuperscript{†} Algoma University, Brampton, Canada}
  \IEEEauthorblockA{\textsuperscript{‡} Bangladesh University of Engineering and Technology (BUET), Dhaka, Bangladesh}
  \IEEEauthorblockA{Emails: \{shadikur, ummekona\}@yorku.ca,\; hasibsourov36@gmail.com,\; syed.danish@algomau.ca}
}
\maketitle

\begin{abstract}

In the digital era, the exponential growth of scientific publications has made it increasingly difficult for researchers to efficiently identify and access relevant work. This paper presents an automated framework for research article classification and recommendation that leverages Natural Language Processing (NLP) techniques and machine learning. Using a large-scale arXiv.org dataset spanning more than three decades, we evaluate multiple feature extraction approaches (TF–IDF, Count Vectorizer, Sentence-BERT, USE, Mirror-BERT) in combination with diverse machine learning classifiers (Logistic Regression, SVM, Naïve Bayes, Random Forest, Gradient Boosted Trees, and k-Nearest Neighbor). Our experiments show that Logistic Regression with TF–IDF consistently yields the best classification performance, achieving an accuracy of 69\%. To complement classification, we incorporate a recommendation module based on cosine similarity of vectorized articles, enabling efficient retrieval of related research papers. The proposed system directly addresses the challenge of information overload in digital libraries and demonstrates a scalable, data-driven solution to support literature discovery.

\end{abstract}


\section{Introduction}

Finding scientific papers or topics can be a challenging yet essential task for researchers, as timely access to the latest knowledge is critical for staying current in one’s field and for conducting high-quality research. Prior studies have highlighted the importance of structured approaches to literature discovery, such as probabilistic topic modeling for uncovering hidden themes in large corpora \cite{griffiths2004finding, blei2003latent}. Traditional strategies for locating scientific papers include querying online databases, exploring institutional repositories, or directly contacting other researchers. However, the exponential growth of digital archives has introduced a significant challenge: information overload. With thousands of articles published daily across platforms like arXiv, PubMed, and IEEE Xplore, it has become increasingly difficult to determine which works are most relevant or impactful. This problem is particularly pronounced for novice researchers or those working in interdisciplinary domains, where the volume and diversity of literature can be overwhelming \cite{jinha2010article}.

Earlier work such as WebFind \cite{monge1996webfind} explored leveraging external sources of information to improve retrieval effectiveness, demonstrating the potential of guided search methods for scientific literature. More recently, advances in natural language processing (NLP) and machine learning (ML) have enabled more sophisticated techniques for text classification, clustering, and recommendation in digital libraries \cite{ikonomakis2005text, dalal2011automatic}. Modern recommender systems, for example, are increasingly applied to academic publishing platforms to support personalized literature discovery, reduce search time, and improve research productivity \cite{beel2016paper}.

Building on these ideas, we propose a novel framework that integrates ML and NLP to automatically classify scientific articles into meaningful categories and recommend related papers. By combining classical text representation techniques (e.g., TF-IDF, Count Vectorizer) with more recent embedding models (e.g., Sentence-BERT, Universal Sentence Encoder), our system addresses both classification accuracy and recommendation relevance. The goal is to reduce the cognitive load of literature search, support efficient navigation of large digital repositories, and improve the discovery of relevant scientific contributions, thereby enhancing the overall research workflow.To structure our investigation, we define the following three research questions:

\vspace{0.5em}
\noindent \textbf{RQ1}: How can the performance of the proposed system be improved for different types of articles?\\
\textbf{Why and How}: We conduct a comprehensive benchmark study to evaluate multiple feature selection and representation techniques for extracting informative features from scientific articles. Feature selection is a crucial aspect of machine learning because it enhances prediction accuracy by retaining relevant attributes while removing noisy or redundant ones. Our evaluation compares a variety of models to identify the most effective techniques for article classification.

\vspace{0.5em}
\noindent \textbf{RQ2}: Which model can better identify scientific articles from the arXiv dataset?\\
\textbf{Why and How}: Using a large-scale dataset of articles from arXiv.org, we experiment with multiple machine learning algorithms and NLP-based feature representations. Specifically, we evaluate six classifiers—Logistic Regression \cite{ho1994decision}, Multinomial Naive Bayes \cite{rennie2003tackling}, Support Vector Machines (SVM) \cite{gunn1998support}, Random Forests \cite{ho1995random}, Gradient Boosted Regression Trees (GBRT) \cite{lanzi2000learning}, and k-Nearest Neighbors (kNN)—to determine the most effective model for classification performance.

\vspace{0.5em}
\noindent \textbf{RQ3}: How can the proposed system be integrated into existing information retrieval or recommendation systems, and what are the potential benefits and challenges of such integration?\\
\textbf{Why and How}: We design a recommendation module based on cosine similarity applied to vectorized article representations. This allows for automatically retrieving related articles in response to user queries. Integrating such a module into existing scholarly search platforms has the potential to enhance efficiency, improve user experience, and expand functionality by delivering more accurate and contextually relevant recommendations. At the same time, we highlight challenges such as scalability, personalization, and generalizability across domains.

\section{Related Work}
\label{sec:rl}
A wide range of studies have investigated article classification and recommendation systems using machine learning (ML) and natural language processing (NLP) techniques. In particular, researchers have extensively applied text mining methods to extract features from unstructured documents, aiming to enhance both classification accuracy and recommendation quality.

Chen et al. \cite{chen2020deep} proposed a deep learning–based framework for patent analysis, employing BiLSTM-CRF and BiGRU-HAN models to identify entities and extract semantic relations. Their results demonstrated that deep neural architectures outperform traditional approaches in capturing semantic structures. Similarly, Rahman et al. \cite{rahman2022example} introduced the Example-Driven Review Explanation (EDRE) method for software engineering, which improves the code review process by recommending relevant examples to clarify feedback. This approach highlights the growing importance of context-aware recommendation in technical domains. Classical approaches to text classification have also been well studied. Ikonomakis et al. \cite{ikonomakis2005text} provided an overview of text classification processes using machine learning techniques, emphasizing their role in information extraction, summarization, retrieval, and question answering. Dalal and Zaveri \cite{dalal2011automatic} reviewed advances in automatic text classification, focusing on challenges such as unstructured text, high-dimensional feature spaces, and the selection of appropriate ML algorithms. These works established the foundation for applying NLP in large-scale digital libraries. Beyond classification, research has also addressed recommendation in scientific literature. Beel et al. \cite{beel2016paper} surveyed academic recommender systems, identifying approaches such as collaborative filtering, content-based filtering, and hybrid methods. They highlighted limitations of simple similarity-based methods and emphasized the need for personalization and contextual understanding. More recently, studies have leveraged deep learning to improve recommendation quality. For instance, Guo et al. \cite{hassan2017personalized} explored neural collaborative filtering and hybrid embedding models for personalized recommendations in scholarly platforms, demonstrating superior performance compared to traditional methods.

Other studies have considered topic modeling and semantic embeddings as enablers of article classification and retrieval. Blei et al. \cite{blei2003latent} introduced Latent Dirichlet Allocation (LDA), a generative probabilistic model widely adopted for uncovering latent topics in scientific corpora. More recently, Reimers and Gurevych \cite{reimers2019sentence} proposed Sentence-BERT, which adapts BERT embeddings for sentence similarity and retrieval tasks, proving effective in recommendation pipelines. Recent studies highlight the broad applicability of NLP and machine learning for text and document analysis. User-generated data has been mined to capture requirements, such as Reddit posts during the COVID-19 pandemic \cite{rahman2023mining}, while CNN-based models have achieved high accuracy in classifying COVID-19 X-rays \cite{basalamah2022optimized}. In media, sentiment analysis of news headlines has been used to detect polarity and reduce misconceptions \cite{rahman2019context}. Topic modeling has also been advanced through improved labeling methods \cite{rahman2020assessing, hossain2019polynomial, rahman2021estimating}. Within software engineering, ML has supported evaluation of code review comments \cite{rahman2025measuring} and improved feedback quality with tools like RefineCode \cite{rahman2024refinecode}. Building on these diverse efforts, our work benchmarks multiple algorithms and embeddings for large-scale scientific article classification and recommendation.


Our study extends prior work by benchmarking ML algorithms and text representation techniques on a large-scale arXiv dataset. Unlike studies that address classification or recommendation separately, our framework integrates both tasks into a unified pipeline. By assessing different algorithm–embedding combinations and their applicability in real-world settings, we provide insights for building scalable systems that help researchers navigate large and growing scientific archives. In particular, our results highlight the value of combining traditional embeddings with modern sentence representations to balance efficiency and accuracy.

\section{Methodology}
\label{sec:meth}
\begin{figure}[b]
\includegraphics[scale=0.43]{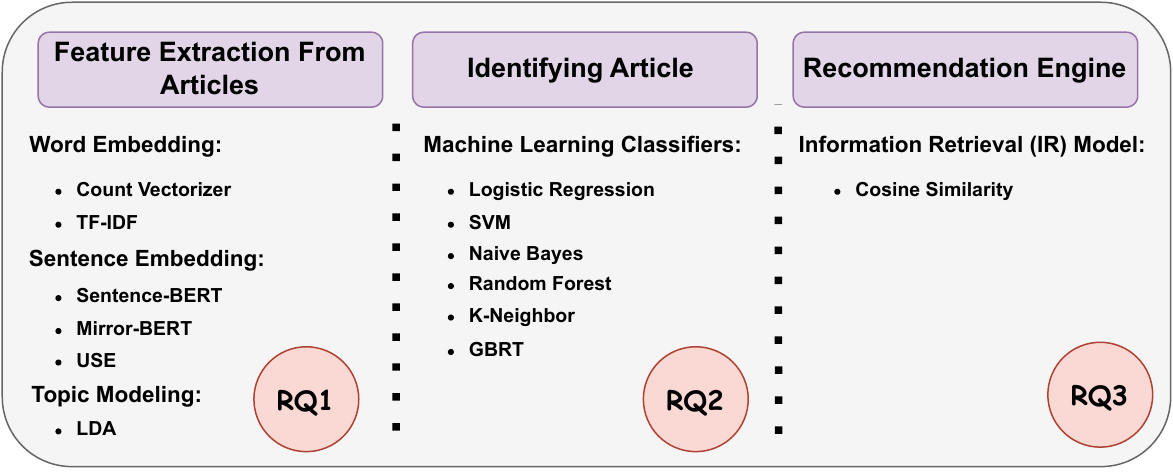}
\caption{Overview of the techniques used for our research methodology.} 
\label{m}
\end{figure}

In this section, we describe the methodological framework employed to design, implement, and evaluate our proposed article classification and recommendation system. The process was structured into four main phases: data gathering, preprocessing, feature extraction, and model training with evaluation, followed by the development of a recommendation engine. Each phase was carefully designed to handle the scale and diversity of the arXiv dataset, ensuring that the raw textual data could be effectively transformed into structured representations suitable for machine learning (ML) and natural language processing (NLP) techniques. Figure~\ref{m} provides an overview of the methodological workflow, highlighting how different components from preprocessing and feature engineering to classification and recommendation were integrated into a unified pipeline.

\subsection{ Data Gathering}
We used the publicly available arXiv dataset\footnote{https://github.com/NeelShah18/arxivData}\footnote{https://www.kaggle.com/datasets/Cornell-University/arxiv}, which contains metadata and abstracts of research articles across multiple disciplines, including Physics, Mathematics, Computer Science, Statistics, Quantitative Biology, Quantitative Finance, Electrical Engineering and Systems Science, and Economics. Each record includes title, abstract, authors, categories, and submission date, with abstracts serving as the primary textual input for classification and recommendation tasks. To prepare the dataset, we removed records with missing abstracts or categories, dropped duplicates, normalized subject labels, lowercased text, removed punctuation, digits, and stopwords, and applied lemmatization, while pruning tokens appearing too rarely or too frequently. Exploratory analysis revealed that Physics, Mathematics, and Computer Science dominate the corpus, accounting for more than 95\% of articles, while fields such as Economics and Quantitative Finance are underrepresented. Abstracts typically range between 120–200 words, with the final cleaned dataset containing approximately 1.2 million records and a vocabulary of ~50,000 unique terms. Submissions span over three decades, showing steady growth, particularly in Computer Science and Machine Learning in recent years.

\subsection{Data Preprossessing}
We have performed various types of text data preprocessing methods to clean our dataset such as tokenization, stop words removal, removing pucntuations, lemmatization etc. With this clean and processed dataset we have performed our study with the help of machine learning as well as NLP methods. Before we could move to data preprocessing, we dropped some fields of the dataset as they are not useful in our analysis. After we have selected the necessary fields for our analysis, we have applied text preprocessing upon our updated dataset.

\textbf{Text Tokenization:} Text tokenization is a splitting method which can splitting operation upon text data and generate smaller parts known as token in the field of NLP. Not only words, pucntuations, numbers from sentence with word tokenization fall into this token category but also sentences from documents with sentence tokenization fall in this category.

\textbf{Removing Punctuations:} This is a process that which is done with the help of regex expressions. We have also performed number removing from our data using this same approach.

\textbf{Removing Stop Words:} Generally, stop words are those words that don't have any 
significant information about the context of the data. These are only used to have grammatical advantages and normally they are removed from the data. Most common stop words are "the","this" "is", "are", "here", "of"  etc.

\textbf{Lemmatization:} Lemmatization process is used to transform words into their lemma or root form e.g. apples-apple, studies-study. This process was really useful to minimize our coupus size as small as possible.

\subsection{Text Mining Process}
After analyzing the preprocessing step, we created the extraction features for the training models based on the corpus text document. Because we need to convert numerical feature vectors to make our text executable on the machine. The machine learning algorithm needs to present text documents in vector form to start training, this process is executed through vector space modeling (VSM). For numerical feature vectors, Bag of Words (TF-IDF method) is a simple techniques. Apart from that, we also consider the Universal Sentence Encoder(USE), Bidirectional Encoder Representations from Transformers (BERT) for sentence embeddings and Mirror-BERT.

\textbf{Count Vectorization:} To our application, we used the count vectorizer for representing the vector. It is a popular and useful technique for text analysis tasks, and can be combined with other techniques such as TF-IDF weighting to improve its effectiveness. This technique involves converting each document into a fixed-length vector representation, where each element of the vector corresponds to the frequency of a particular word or term in the document. Count Vectorization is simple and efficient, and can be applied to a large number of documents with minimal computational overhead.

 \textbf{Term Frequency-Inverse Document Frequency (TF-IDF):} In our study, we used the TF-IDF vectorizer for representing the vector. TF-IDF is a popular text vectorization technique used in natural language processing to analyze and prepare textual data. The approach works by weighting the importance of words in a document based on the frequency of their occurrence in the corpus of documents. It is an extension of the bag-of-words model, which is a simple representation of text that ignores the order of words and focuses on their occurrence. TF-IDF vectorization assigns higher weight to words that are more important in a document and less frequent in the corpus. This means that common words such as "the" and "a" are assigned a lower weight while rare words are given higher importance. On the other hand, we also used the N-Gram process for sequences between words. TF-IDF vectorization is a useful technique for several NLP tasks, including text classification, clustering, and information retrieval.

\textbf{Sentence-BERT:} To consider the bidirectional representation of the words we also invested in using the BERT  model. Bidirectional Encoder Representations from Transformers (BERT)\cite{devlin2018bert} is a transformer-based open-source machine learning framework for natural language processing (NLP). While we used TF-IDF to extract the term frequency (which works at the word level) we use the BERT model to consider the bidirectional representation of words at the sentence level. For that reason, we consider the Sentence-BERT\cite{reimers2019sentence} for the sentence embeddings. Sentence-BERT is a sentence embedding technique that utilizes the BERT network. It uses a siamese network architecture to enable the generation of high-quality sentence embedding. Sentence-BERT requires human-annotated labeled data for fine-tuning the sentence embeddings.

\textbf{Universal Sentence Encoder:} Universal Sentence Encoder(USE) is another sentence embedding technique by Google Research\cite{cer2018universal}. For text mining, we also consider the Universal Sentence Encoder(USE) for sentence embeddings.  Universal Sentence Encoder (USE) is a sentence embedding technique by Google Research. USE has two separate ways of encoding sentences. The first USE model is based on a simple transformer encoder block. This model constructs an encoding subgraph of transformer architecture to generate sentence embeddings. The subgraph utilizes attention to capture the context as well as the sequence of words to generate high-quality embeddings. In this method, the sentence embeddings are generated from word embeddings by simply computing an element-wise sum of the individual word representations. Further, multi-task learning is used to fine-tune sentence embeddings. The tasks involved in the multi-task learning include: Skip-Thought \cite{kiros2015skip}, Conversation task \cite{henderson2017aszl} and various supervised tasks of text classification. The second variant of USE also utilizes the transformer encoder attention block however, there are some modifications done to the overall architecture as compared to its first variant. The second method is called Deep Averaging Network (DAN).

\textbf{Mirror-BERT:} Mirror-Bert~\cite{liu2021fast} avoids the need for external human-labeled data for turning MLMs into sentence encoders by utilizing contrastive learning. The idea behind contrastive learning is to attract the text belonging to the same class (i.e. positives) while pushing away the negatives. It is extremely easy to generate negative examples. Negative examples can be simply generated by random sampling. Positive data is sometimes created manually however, Mirror-BERT automates the process of creating positive pairs by introducing two corruption techniques. The underlying assumption here is that by applying these so-called corruptions, the original meaning of the sentence will still remain intact and thus, the original representation and the corrupted representation should be very similar.

{\textbf{Latent Dirichlet allocation(LDA)}}
Latent Dirichlet Analysis is a probabilistic model, and to obtain cluster assignments. LDA\cite{blei2003latent} utilizes Dirichlet priors for the document-topic and word-topic distributions and it utilizes two probability values: P (word - topics) and P (topics - documents).To perform the LDA model works by deﬁning the number of ’topics’ that are begun in our set of training text documents. Now we will present the model output below: here we have chosen the number of topics = 20 and Number of Words = 20. We have also set the random state = 1, which is enough for our journals and articles' text documents.

\subsection{Training Model}
To build a robust article classification framework, we adopted a supervised learning approach based on the \texttt{OneVsRestClassifier}, which is well-suited for handling multi-label classification tasks. In this setting, each research article can be associated with one or more categories (e.g., computer science, mathematics, physics), requiring algorithms that can decompose the problem into multiple binary classification subtasks. The \texttt{OneVsRestClassifier} achieves this by training a separate classifier for each class, where the classifier predicts whether the instance belongs to the corresponding category. 

For the underlying classifiers, we experimented with four widely used algorithms: Logistic Regression, Multinomial Naïve Bayes (MNB), Linear Support Vector Classifier (SVC), and Random Forest. Each of these models captures different inductive biases: Logistic Regression provides a strong linear baseline; MNB is particularly effective for text classification with sparse data; Linear SVC is robust in high-dimensional spaces; and Random Forest leverages ensemble learning to capture nonlinear decision boundaries. By combining these algorithms with the \texttt{OneVsRestClassifier}, we sought to identify the most effective approach for our corpus.

\begin{figure*}[h]  
  \centering
\includegraphics[width=\textwidth,height=\textheight,keepaspectratio]{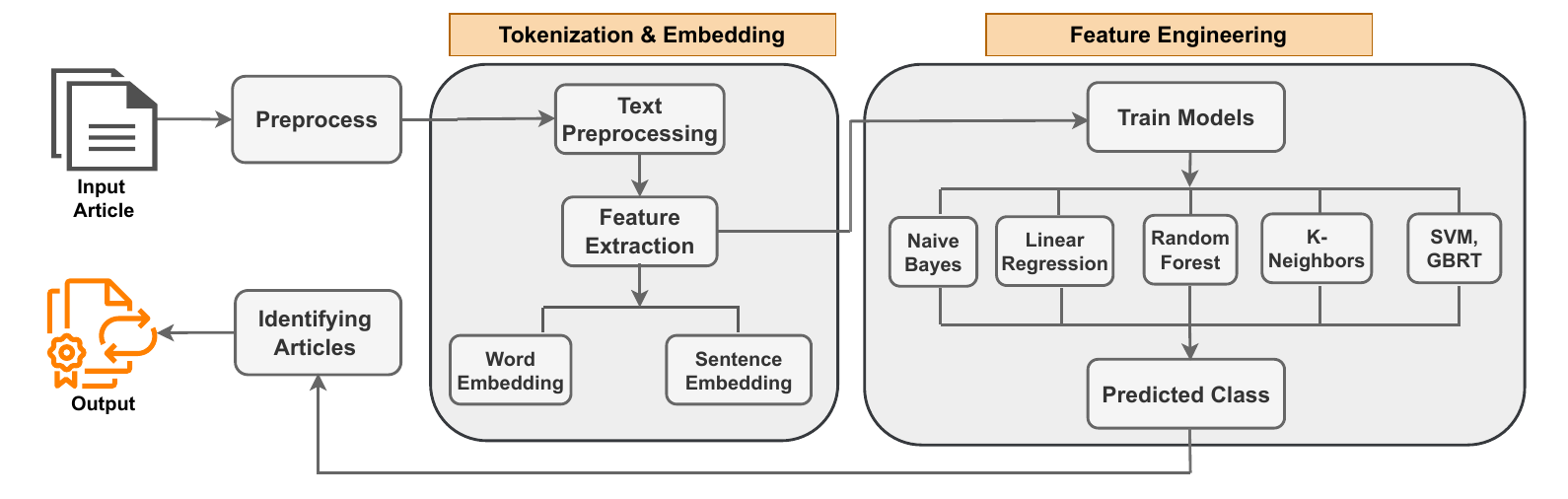}
  \caption{Comprehensive overview of the training pipeline for article classification. The process begins with preprocessing steps followed by feature engineering. These feature vectors are then used to train multiple classifiers, enabling the system to predict article categories with improved accuracy and generalization.}
  \label{fig:train}
\end{figure*}

The training process, illustrated in Figure~\ref{fig:train}, begins with preprocessing and vectorization of the arXiv dataset, followed by model training and evaluation. To ensure optimal performance, we employed \texttt{GridSearchCV} from Scikit-learn for hyperparameter optimization, including regularization strength for Logistic Regression, smoothing parameter $\alpha$ for Naïve Bayes, kernel/penalty terms for SVC, and tree depth/number of estimators for Random Forest. Cross-validation with $n=10$ folds was used to evaluate different parameter combinations, mitigating the risk of overfitting and ensuring generalization.

For model validation, we followed a standard 80/20 split, using 80\% of the dataset for training and reserving the remaining 20\% for testing. This split provides a large enough sample for model fitting while maintaining a sufficient holdout set for unbiased evaluation. The training set was used to fit the classifiers, while the test set served as a proxy for unseen data to estimate generalization performance. Evaluation metrics included accuracy, precision, recall, and F1-score, which allowed us to capture both the overall correctness of predictions and the balance between false positives and false negatives. 

Through this systematic comparison of algorithms and vectorization methods, our goal was to determine the best-performing pipeline for scientific article classification. The insights gained from this benchmarking not only guided the choice of classifier but also informed how different models trade off interpretability, efficiency, and predictive accuracy in the context of large-scale textual datasets such as arXiv.

\subsection{Recommendation System}  

To enhance user experience and support efficient literature discovery, we designed a recommendation system that suggests research articles most relevant to a user’s interests. Our approach is based on \textbf{cosine similarity}, a widely used technique in information retrieval and natural language processing for measuring the closeness of documents in a high-dimensional vector space~\cite{singhal2001modern}. The central idea behind cosine similarity is that two documents are considered similar if their vector representations point in the same direction, regardless of their absolute magnitude. Mathematically, cosine similarity between two vectors $\mathbf{A}$ and $\mathbf{B}$ is defined as:  

\begin{equation}
\text{Cosine Similarity}(\mathbf{A}, \mathbf{B}) = 
\frac{\mathbf{A} \cdot \mathbf{B}}{\|\mathbf{A}\| \|\mathbf{B}\|}
\end{equation}

where the numerator denotes the dot product of the two vectors, and the denominator normalizes their magnitudes. This normalization makes the measure scale-invariant, allowing long documents and short documents to be compared fairly.  

In our system, each research article is first \textbf{vectorized} using feature extraction techniques described in the previous subsection, such as TF--IDF, Count Vectorizer, or semantic embeddings (e.g., Sentence-BERT, Universal Sentence Encoder). These representations transform unstructured abstracts into fixed-length vectors that capture both lexical and semantic characteristics. Once the corpus is represented in this form, cosine similarity is applied to compute pairwise similarity scores between articles.  

To generate recommendations, the system takes into account a user’s \textbf{reading history} or \textbf{search queries}. For each document in the user’s profile, the engine retrieves the top-$k$ most similar articles by ranking all candidate documents according to their cosine similarity scores. These results are then aggregated to produce a \textbf{personalized recommendation list}, which highlights articles that are most contextually aligned with the user’s current research focus. This approach ensures that recommendations are not only accurate in terms of topical similarity but also \textbf{dynamic}, adapting as a user’s reading history evolves over time.  

While our implementation is lightweight and computationally efficient, it remains highly effective for large-scale repositories such as \textit{arXiv}, where millions of research articles are continuously added. Moreover, the method is easily extensible: additional filters (e.g., publication date, category, citation count) can be incorporated to refine results and improve user satisfaction. Overall, our recommendation system leverages the strength of cosine similarity to bridge the gap between massive academic datasets and individual researchers’ needs, providing a scalable and interpretable solution for scholarly discovery.

\section{Results}
\label{sec:result}
In this section, we consider the overall results of \textbf{RQ1}, \textbf{RQ2} and \textbf{RQ3} and further discuss the implications of our research work. we evaluate the text classification and recommendation performance of our models based on the research articles. In this results section, we consider the overall results of our research work. To evaluate the text classification performance, we collected research articles to perform our text classification results. After that, we built the text classifier with machine learning by implementing several classifiers to classify research articles ( Machine Learning, Artificial Intelligence, Deep Learning, Computer Vision, Computer Science, Software Engineering etc. ). Apart from that, we built the recommendation system based on articles. To develop a recommender application tool, we have performed the cosine similarity method based on articles. 

\textbf{RQ1: How can the performance of the proposed system be improved for different types of articles?}

To investigate this question, we conducted a comprehensive benchmark study to identify the most effective feature representations for training classification models on article summaries. We compared multiple text representation approaches, including traditional word-based embeddings such as Term Frequency–Inverse Document Frequency (TF–IDF) and Count Vectorizer, as well as sentence-level embeddings generated by Mirror-BERT, Sentence-BERT, and the Universal Sentence Encoder (USE). These methods were applied to capture semantic and syntactic information from the textual data and to assess their impact on classification performance.

In addition to embedding-based features, we employed topic modeling to enrich the feature space with latent semantic information. Specifically, we used Latent Dirichlet Allocation (LDA) implemented via both the Gensim library and the Mallet toolkit. LDA enabled us to capture high-level thematic structures within the articles, generating topic distributions that we appended as an additional feature column to the classification dataset. This integration allowed the models to leverage both direct textual embeddings and inferred topical information.

The preprocessing pipeline included tokenization, punctuation removal, stop-word elimination, POS tagging, and lemmatization. To enhance semantic consistency, we also constructed bigram and trigram models, followed by dictionary (id2word) and corpus generation restricted to specific part-of-speech categories (nouns, adjectives, verbs, and adverbs). These steps ensured a focused and structured representation of the text for downstream analysis.

\begin{figure}[h]
\centering
\includegraphics[scale=0.65]{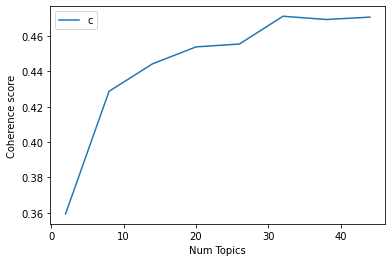}
\caption{Optimal number of topics for LDA based on coherence scores.}
\label{lda_cohe}
\end{figure}

We evaluated the quality of the LDA models using perplexity and coherence scores. The initial results with standard LDA produced a perplexity of $-7.63$ and a coherence score of $0.39$, which indicated limited interpretability. To address this, we adopted the Mallet implementation of LDA, which is known for producing more coherent topics. As shown in Figure~\ref{lda_cohe}, the coherence-based evaluation helped determine the optimal number of topics for feature generation.

\begin{tcolorbox}[colback=gray!10, colframe=black, boxrule=0.3mm, arc=0mm, left=2mm, right=2mm, top=1mm, bottom=1mm]
Among the various embedding methods, \texttt{TF--IDF} achieved the most consistent performance for article classification. Furthermore, incorporating topic distributions from LDA as an additional feature column enhanced the richness of the dataset and provided complementary semantic information for model training.
\end{tcolorbox}


\textbf{RQ2: Which model can better identify scientific articles from the arXiv dataset?}

To classify the research articles, we have trained various types of machine learning algorithms and compared them to find the best classifier from the results. Considering Mirrot-BERT, we used six machine learning classifiers. Among those, the Logistic Regression, Support Vector Machine (SVM) and Naive Bayes classifier achieve higher accuracy (0.52) than other classifiers. The table \ref{Sentence} shows the accuracy of our training models.

\begin{table*}[h]
\centering
\setlength{\tabcolsep}{2.9pt} 
\caption{Result of classifiers trained on various Sentence embedding models}\label{Sentence}
\begin{tabular}{|l | c c c c|c c c c|c c c c|}
\hline
\rowcolor{lightgray}MODEL & \multicolumn{4}{c}{Mirror-BERT} & \multicolumn{4}{c}{Sentence-BERT} & \multicolumn{4}{c}{USE} \\
\hline
\hline
\textbf{Classifiers} &\textbf{Precision} & \textbf{Recall} & \textbf{F-Score} & \textbf{Accuracy} & \textbf{Precision} & \textbf{Recall} & \textbf{F-Score} &  \textbf{Accuracy} & \textbf{Precision} & \textbf{Recall} & \textbf{F-Score} & \textbf{Accuracy}\\
\hline
Logistic Regression & 0.62 & 0.51 & 0.52 & \textit{0.52*} & 0.65 & 0.53 & 0.55 & 0.53  & 0.63 & 0.52 & 0.53 & 0.51 \\
SVM & 0.65 & 0.51 & 0.52 & 0.52 & 0.68 & \textit{0.54} & \textit{0.53} & \textit{0.55*} & 0.67 & 0.51 & \textit{0.51} & \textit{0.53*} \\
Naive Bayes & 0.75 & 0.49 & 0.53 & 0.52 & 0.71 & 0.51 & 0.56 & 0.52 & 0.70 & 0.54 & 0.56 & 0.50 \\
Random Forest & 0.64 & 0.48 & \textit{0.54*} & 0.50 & 0.74 & 0.58 & \textit{0.57*} & 0.51 & 0.72 & 0.56 & \textit{0.58*} & \textit{0.52}\\
K-neighbour & 0.59 & 0.53 & 0.51 & 0.50 & 0.60 & 0.58 & 0.52 & 0.51 & 0.62 & 0.58 & \textit{0.50} & \textit{0.49}\\
GBRT & 0.53 & 0.49 & 0.52 & 0.48 & 0.58 & 0.50 & 0.49 & 0.49 & 0.57 & 0.50 & 0.49 & 0.49\\
\hline
\end{tabular}
\end{table*}

\begin{table}[h]
\centering
\setlength{\tabcolsep}{1pt} 
\caption{Result of classifiers trained on word embedding models}\label{Word}
\begin{tabular}{|l | c c c c|c c c c|}
\hline
\rowcolor{lightgray}MODEL & \multicolumn{4}{c}{TF-IDF} & \multicolumn{4}{c}{Count-Vectorizer}  \\
\hline
\hline
\textbf{Classifiers} &\textbf{Precision} & \textbf{Recall} & \textbf{F} & \textbf{Acc.} & \textbf{Precision} & \textbf{Recall} & \textbf{F} &  \textbf{Acc.}  \\
\hline
Logistic Reg. & 0.82 & 0.51 & 0.63 & \textit{0.69*} & 0.72 & 0.50 & 0.53 & \textit{0.67*} \\
SVM & 0.77 & 0.56 & \textit{0.65*} & 0.65 & 0.75 & 0.53 & 0.55 & 0.60 \\
Naive Bayes & 0.80 & 0.50 & 0.62 & 0.65 & 0.65 & 0.50 & 0.52 & 0.63  \\
Random Forest & 0.84 & 0.54 & 0.59 & 0.64 & 0.74 & 0.52 & \textit{0.56*} & 0.62 \\
K-neighbour & 0.71 & 0.53 & 0.58 & 0.53 & 0.69 & 0.52 & \textit{0.56*} & 0.52 \\
GBRT & 0.70 & 0.54 & 0.59 & 0.53 & 0.63 & 0.50 & 0.53 & 0.51 \\
\hline
\end{tabular}
\end{table}

Using the tf-idf, we have performed multiple machine learning classifiers for text classification based on articles. Among those, the Logistic Regression classifier achieves higher accuracy (0.69) than other classifiers Support Vector Machine, Naive Bayes, Random Forest, K-neighbour, and Gradient Boosted Regression Trees (GBRT) (0.65, 0.65, 0.64, 0.53 and 0.53 respectively). The table  ~\ref{Word} shows the result of multiple machine learning classifiers trained on tf-idf.

We also used count-vectorizer to perform multiple machine learning classifiers for text classification based on articles. Among those, the Logistic Regression classifier achieves higher accuracy (0.67) than other classifiers Support Vector Machine, Naive Bayes, Random Forest, K-neighbour, and Gradient Boosted Regression Trees (GBRT) (0.60, 0.63, 0.62, 0.52 and 0.51 respectively). The table  ~\ref{Word} shows the result of multiple machine learning classifiers trained on tf-idf.

\begin{figure*}[t]
\includegraphics[scale=0.67]{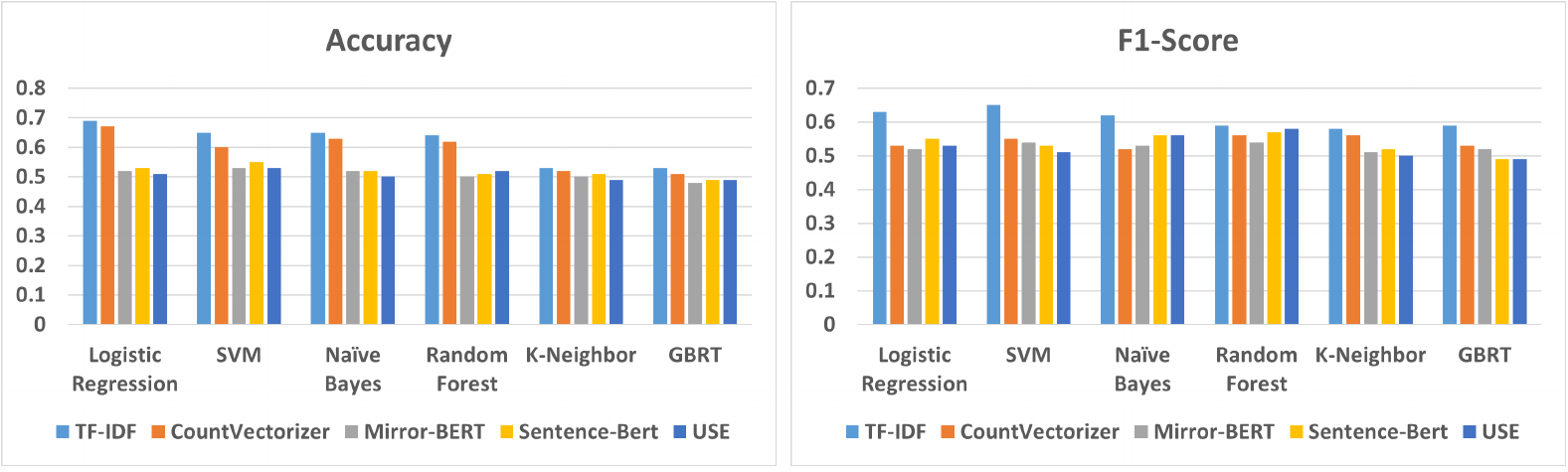}
\caption{Comparison of models performance based on (a) Accuracy and (b) F-score. Logistic Regression trained on word embedding (TF-IDF) performs better in both metrics.} 
\label{visualcompare}
\end{figure*}

we also considered the Sentence-BERT encoder for sentence embedding. For that reason, we used several machine learning classifiers for text classification based on research articles. Among those, the Support Vector Machine classifier achieves higher accuracy (0.55) than other classifiers Random Forest, Logistic Regression, Naive Bayes, K-neighbour and Gradient Boosted Regression Trees (GBRT) (0.51, 0.53, 0.52, 0.51 and 0.49 respectively). The table \ref{Sentence} shows the result of multiple machine learning classifiers trained on Sentence-BERT.

Apart from that, we have considered the universal sentence encoder(USE) for sentence embedding. For that reason, we applied various machine learning classifiers for the text classifying based on code review comments. Among those, the Support Vector Machine (SVM) classifier achieves higher accuracy (0.53) than other classifiers Random Forest, Logistic Regression,  Naive Bayes, k-neighbour, and Gradient Boosted Regression Trees (GBRT) (0.52, 0.51, 0.50, 0.49 and 0.49 respectively). The table \ref{Sentence} shows the result of several machine learning classifiers trained on universal sentence encoder(USE).

\begin{tcolorbox}[colback=gray!10, colframe=black, boxrule=0.3mm, arc=0mm, left=2mm, right=2mm, top=1mm, bottom=1mm]
Logistic Regression model, trained with a word embedding technique, demonstrated the highest level of accuracy in predicting and identifying specific research articles, achieving an F-score of 0.69 for researchers.
\end{tcolorbox}

\textbf{RQ3: How can the proposed system be integrated into existing information retrieval or recommendation systems, and what are the potential benefits and challenges of such integration?}

The recommendation of research articles plays a crucial role in assisting researchers with their work. With the ever-increasing volume of research articles available online, it can be challenging for researchers to keep up with the latest findings and identify relevant content. Recommendations can help researchers to navigate large archives and discover new research that is related to their interests and work. By automating the process of recommendation through the use of Natural Language Processing (NLP) techniques and machine learning algorithms, researchers can save time and effort in searching for articles, while also ensuring that the recommendations are accurate and relevant. Furthermore, the use of personalized recommendation systems can tailor the recommendations to the individual researcher's interests and preferences, providing an even more effective tool for assisting them in their work. In order to provide researchers with a tool that can help them find relevant research articles based on their specific needs, we applied Cosine Similarity as an Information Retrieval (IR) model. This approach allows us to compute the similarity between research articles and search queries, enabling us to provide researchers with the top twenty most similar articles. By leveraging the power of IR and cosine similarity, we can quickly and accurately identify research articles that are most relevant to a given search query. This tool is particularly useful for researchers who are exploring a new topic or who need to quickly find relevant research to support their work. The ability to quickly and accurately identify relevant research articles can save researchers time and effort, while also ensuring that they have access to the most up-to-date and relevant information. Overall, the application of Cosine Similarity as an IR model offers a powerful tool for assisting researchers in their work and advancing the state of knowledge in their respective fields.

\begin{tcolorbox}[colback=gray!10, colframe=black, boxrule=0.3mm, arc=0mm, left=2mm, right=2mm, top=1mm, bottom=1mm]
Our application tool utilizes cosine similarity between research articles to generate a meaningful list of articles, offering relevant suggestions of similar research articles to the researcher.
\end{tcolorbox}

\section{Threats to Validity}
There are several potential threats to the validity of our research based on the methods and techniques we employed. Firstly, the use of a public dataset collected from archive.org may limit the generalizability of our findings to other datasets. Secondly, while we used multilabel classification and several NLP and machine learning techniques, the choice of these techniques and their parameters could introduce bias into our results. Additionally, the data labels were not balanced, which could have affected the accuracy of our classification model. Although we used SMOTE to balance the data, this technique may not have been fully effective in addressing the imbalance. Furthermore, despite using 10-fold cross-validation, the accuracy of our classification model was still only in the range of 50 to 65 percent, which may indicate that our data was unstructured and noisy. Finally, the limitations of our methodology, including the use of specific algorithms and techniques, may restrict the generalizability of our findings to other research domains.

\section{Conclusion and Future Work}
\label{Conclusion}

The classification and recommendation of research articles remains a critical task in the digital era, as the exponential growth of online repositories has made it increasingly difficult for researchers to identify relevant content efficiently. In this work, we presented a machine learning–based framework that leverages Natural Language Processing (NLP) techniques to automatically categorize and recommend articles. Our approach not only streamlines literature discovery but also provides valuable guidance for novice researchers who may be unfamiliar with a particular domain, enabling them to navigate large archives and uncover relevant research more effectively. The results of our study demonstrate the potential of combining traditional embeddings with advanced sentence representations to achieve both accuracy and scalability in article classification and recommendation.

While the results demonstrate the promise of our approach, there is significant room for further advancement. As future work, we plan to evaluate the system on larger and more diverse datasets to assess its generalizability across domains. We also aim to investigate the integration of deep learning models, such as transformer-based architectures, for improved classification and recommendation performance. Another direction will be the incorporation of user feedback and adaptive learning mechanisms to refine recommendations over time. Finally, we intend to design and implement an intuitive user interface, making the system more accessible and practical for researchers with varying levels of technical expertise.

\bibliographystyle{IEEEtran}
\bibliography{article}
\end{document}